\documentclass[prd,twocolumn,showpacs,preprintnumbers,aps,nofootinbib]{revtex4-1}
\usepackage{graphicx}
\usepackage{dcolumn}
\usepackage{bm}
\usepackage{amsmath,amssymb,amsfonts}
\usepackage{latexsym}
\usepackage{color}

\begin{document}


\title{\boldmath
Probing for Extra Top Yukawa Couplings in Light of $t\bar th(125)$ Observation
 }

\author{Wei-Shu Hou, Masaya Kohda and Tanmoy Modak}
\affiliation{Department of Physics, National Taiwan University, Taipei 10617, Taiwan}
\bigskip

\date{\today}

\begin{abstract}
The observation of $t\bar th(125)$ production at the Large Hadron Collider (LHC)
is the first direct measurement of the top Yukawa coupling.
It opens the window on an extra top Yukawa coupling, $\rho_{tt}$,
from a second Higgs doublet, without a $Z_2$ symmetry
to forbid flavor changing neutral Higgs couplings.
We show that $t\bar th$ and Higgs property measurements at
the High Luminosity LHC can constrain
the ${\rm Re}\,\rho_{tt}$--${\rm Im}\,\rho_{tt}$ parameter space
that could drive electroweak baryogenesis,
but the $\Gamma_h$ width measurement must be
considerably improved beyond current projections.
\end{abstract}


\maketitle

\section{Introduction} 

The CMS and ATLAS experiments announced~\cite{Sirunyan:2018hoz, Aaboud:2018urx}
recently the observation of $pp \to t\bar th^0$ production
at the Large Hadron Collider (LHC),
where $h^0$ is the 125 GeV scalar boson discovered
not so long ago. The observed production strengths
are consistent with the Standard Model (SM) expectation,
and constitute the landmark direct detection of
the Higgs boson coupling to the top quark,
the top Yukawa coupling $\lambda_t$.
While indirect evidence have long existed, we now have
direct experimental proof that $\lambda_t \simeq 1$,
as prescribed by $\lambda_t = \sqrt2 m_t/v$ in the SM,
where $v$ is the vacuum expectation value of
the Higgs doublet field.

Let us recap the experimental observations.
Using 35.9 fb$^{-1}$ data collected in 2016 at 13 TeV
collision energy and covering the
$h \to WW^*$, $ZZ^*$, $\gamma\gamma$, $\tau^+\tau^-$
and $b\bar b$ final states, together with Run 1 data
taken at 7 and 8 TeV,
CMS~\cite{Sirunyan:2018hoz} observes the relative strength
\begin{align}
 \mu_{t\bar th} = 1.26^{+0.31}_{-0.26},
   \quad \ \, ({\rm Run\; 2}+ {\rm Run\;1,\; CMS})
\label{eq:mu-CMS}
\end{align}
with respect to SM, amounting to 5.2$\sigma$ significance.
The expected significance for SM is 4.2$\sigma$.
ATLAS had published earlier~\cite{Aaboud:2017jvq}
a Run 2 result based on 36.1 fb$^{-1}$ collected in 2016
and covering the same Higgs decay final states,
finding 4.2$\sigma$ evidence.
As a contribution to the LHCP conference held recently
in Bologna, ATLAS updated the $h \to \gamma\gamma$ and
$h \to ZZ^* \to 4\ell$ modes to a total of 79.8 fb$^{-1}$
data at 13 TeV. Together with the other 3 modes
based on 36.1 fb$^{-1}$, ATLAS observes~\cite{Aaboud:2018urx}
\begin{align}
 \mu_{t\bar th} = 1.32^{+0.28}_{-0.26},
   \quad \ \, ({\rm Run\; 2\; Update,\; ATLAS})
\label{eq:mu-ATL}
\end{align}
%
at 5.8$\sigma$ significance (4.9$\sigma$),
while combining further with Run 1 data,
the significance becomes 6.3$\sigma$ (5.1$\sigma$).
We have put the SM expectation in parenthesis.

Thus, both ATLAS and CMS have \emph{observed} $t\bar th$ production.
We note that both experiments had
earlier hints for $t\bar th$ production with
strength stronger than SM, which improve
the combined significance quoted above.

Since Yukawa couplings are the source of $CP$ violation (CPV) in SM,
with direct measurement of top Yukawa coupling attained,
it opens up the question whether there are
\emph{additional} Yukawa couplings.
As there is nothing against the existence of a second doublet,
the two Higgs doublet model (2HDM) is in fact one of
the most plausible beyond-SM (BSM) possibilities, where one
should ``naturally'' have a second set of Yukawa couplings.
However, due to the perceived curse of
flavor changing neutral Higgs (FCNH) couplings,
Glashow and Weinberg famously
eliminated all such extra couplings by demanding
``natural'' flavor conservation (NFC)~\cite{Glashow:1976nt}:
each type of fermion charge receives mass from
{\it just one} Higgs doublet.
This enforces only one Yukawa matrix per
fermion mass matrix, and they are therefore
simultaneously diagonalized: absence of FCNH.
There are two ways to implement the NFC condition for quarks,
and 2HDM-II, where $u$-type and $d$-type quarks
receive mass from separate doublets, has been
the most popular, as it arises with supersymmetry.

A decade after the Glashow-Weinberg
NFC condition, the emerging quark mass-mixing hierarchy
led to the critique~\cite{Cheng:1987rs} that NFC may be overkill.
As the top quark is the heaviest fermion,
the best probe may be $t \to ch$~\cite{Hou:1991un}
for FCNH $tch$ coupling.
With the observation of $h(125)$ in 2012, it was stressed~\cite{Chen:2013qta}
that the $2\times 2$ extra Yukawa couplings $\rho_{cc},\; \rho_{ct},\; \rho_{tc}$,
and $\rho_{tt}$ of the \emph{exotic} doublet should be taken 
seriously~\cite{rhoui},
and the issue is experimental: we have to demonstrate their nonexistence,
rather than \emph{assume} NFC and throw them away.
In this paper we address the importance of the
flavor-diagonal $\rho_{tt}$.

\begin{figure*}[t!]
\center
 \includegraphics[width=6cm]{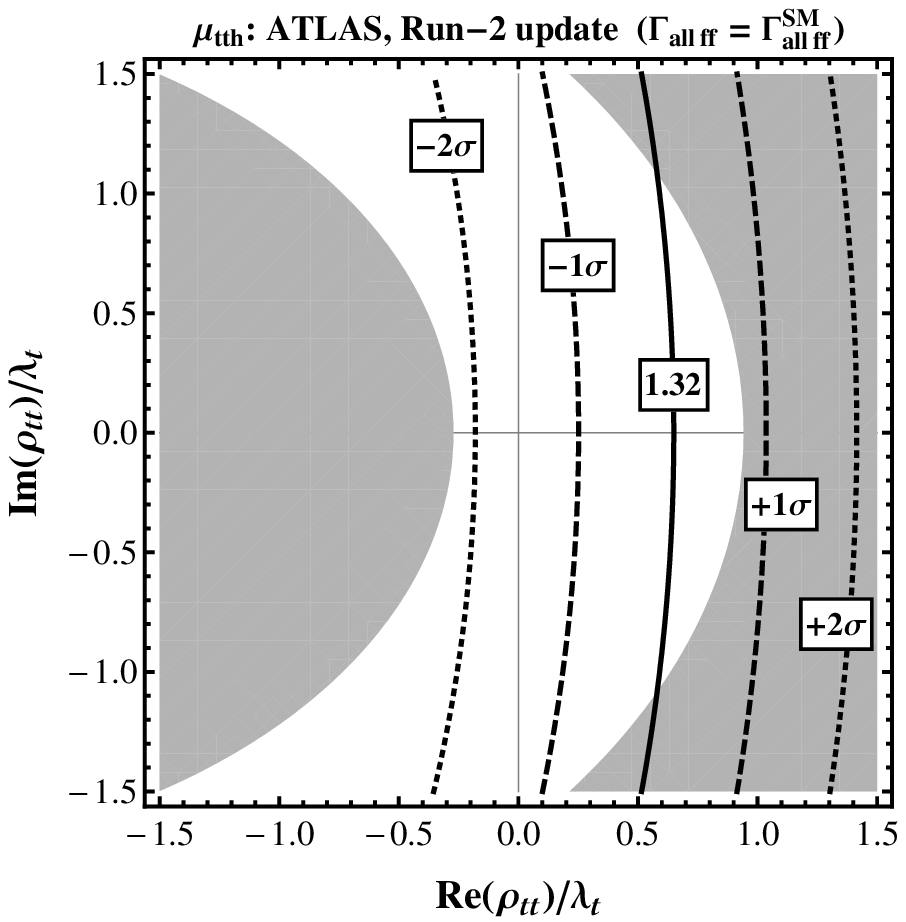} \hskip0.3cm
 \includegraphics[width=6cm]{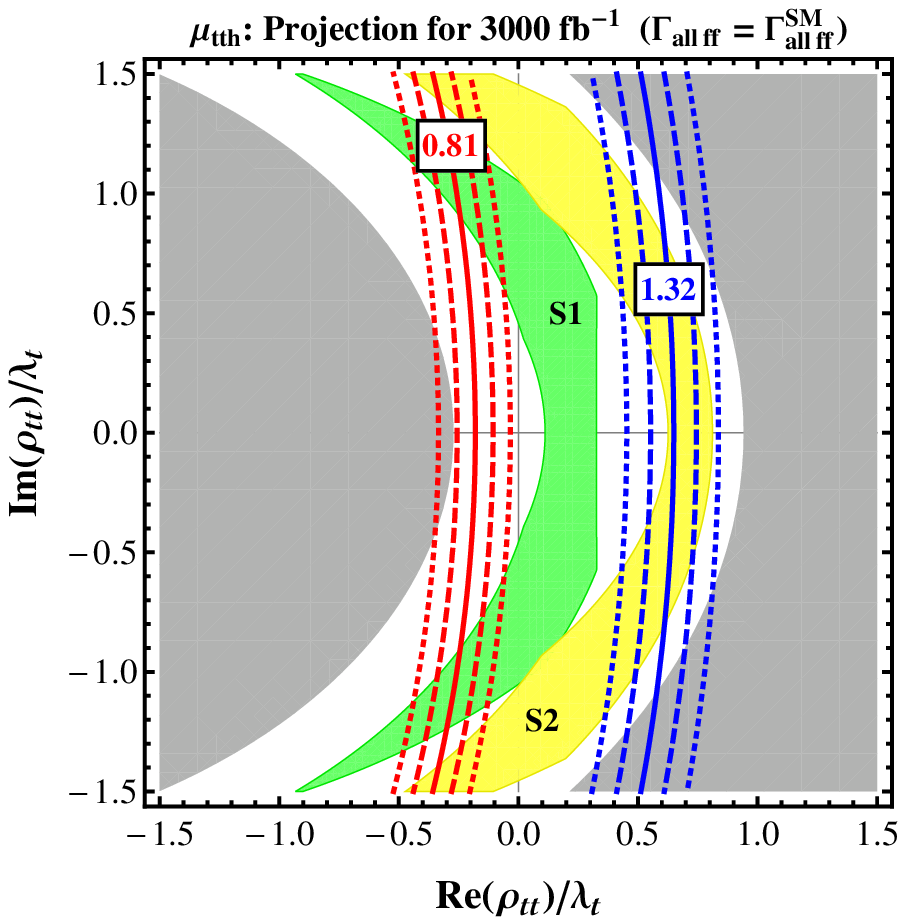}
\caption{
In both panels, the gray regions represent $2\sigma$ exclusion by Run 
1 ATLAS-CMS combination~\cite{Khachatryan:2016vau}, while $\mu_{t\bar t h}$ (solid) with dashed 
(dotted) lines for $1\sigma$ ($2\sigma$) range are for [Left] ATLAS Run-2 update~\cite{Aaboud:2018urx}; 
[Right] projection to full HL-LHC data with central values as marked, allowing 5\% and 10\% uncertainties.
The green (yellow) region in the right panel gives 
the projected $2\sigma$ allowed region in Scenario 1 (2), 
for full HL-LHC data defined in Eq.~(\ref{eq:ggF-proj}).
We have taken $\cos\gamma = 0.3$ and $\sin\gamma < 0$.
}
\label{fig:fig1}
\end{figure*}

The $tth$ coupling in 2HDM without NFC is
\begin{align}
\mathcal{L}_{tth}
= -\frac{1}{\sqrt{2}}\bar t_L (-\lambda_t \sin\gamma +\rho_{tt} \cos\gamma) t_R h
 +{\rm h.c.},
\label{eq:tth}
\end{align}
while the $tch$ coupling is ($\rho_{ct}$ is already constrained
by flavor physics to be small~\cite{Chen:2013qta, Altunkaynak:2015twa})
\begin{align}
\mathcal{L}_{tch}
= -\frac{1}{\sqrt{2}}\rho_{tc}\cos\gamma~\bar t_L c_R h,
\label{eq:tch}
\end{align}
where $\cos^2\gamma \ll 1$ is the alignment phenomenon
observed at the LHC~\cite{cos_b-a}, the fact that the observed
$h$ boson is rather close~\cite{Khachatryan:2016vau} to the SM Higgs boson.
In 2HDM-II, the mixing angle of the two $CP$-even scalars,
$\cos\gamma$,
is usually expressed as $\cos(\alpha -\beta)$,
but without the NFC condition, or the $Z_2$ symmetry to implement it,
$\tan\beta$ is unphysical, hence we use the different notation~\cite{Hou:2017hiw}.
It was recently noted that $\lambda_t {\rm Im}\,\rho_{tt}$
can easily drive~\cite{Fuyuto:2017ewj}
the baryon asymmetry of our Universe (BAU),
or electroweak baryogenesis (EWBG),
which needs ${\cal O}(1)$ Higgs quartic couplings of the 2HDM
for the first order electroweak phase transition.
The latter can relatively easily accommodate~\cite{Hou:2017hiw}
the observed approximate alignment phenomenon.
These add to the attraction of 2HDM 
without the NFC condition.
Note that EWBG is not suppressed by $\cos\gamma$~\cite{Fuyuto:2017ewj}.
Other processes not suppressed by $\cos\gamma$ generally
involve the exotic Higgs bosons directly,
such as $b\to s\gamma$~\cite{Chen:2013qta, Altunkaynak:2015twa},
$gg \to H^0/A^0 \to t\bar t,\; t\bar c$~\cite{Altunkaynak:2015twa},
and $cg \to tH^0/tA^0 \to tt\bar t,\; tt\bar c,\; t\bar tc$~\cite{Kohda:2017fkn}.

So far, the $t\to ch$ decay has not been observed,
with limits approaching $10^{-3}$~\cite{tch-13}.
We assume it is suppressed by $\rho_{tc}$,
but take the maximum $\cos\gamma$ value allowed by data.
The point is, even if the admixture of $\rho_{tt}$
into the $tth$ coupling is suppressed by approximate alignment,
i.e. $\cos\gamma$, interference with the leading SM $\lambda_t$ effect
provides a sensitive probe in $t\bar th$ production.
In the following, we illustrate the new,
direct probe of $t\bar th$ production, and compare with
indirect probes of $h \to gg$, $\gamma\gamma$ loop processes,
projecting into the future of High-Luminosity LHC (HL-LHC).
We give a simplified discussion of how other extra Yukawa couplings,
such as $\rho_{bb}$, would affect $\Gamma_h$
and shift the loop constraint.
We offer some remarks on future prospects beyond HL-LHC.

\section{\boldmath Indirect and Direct Probes of $\rho_{tt}$} 

$\rho_{tt}$ and $\cos\gamma$ are constrained by ATLAS and CMS measurements
of the Higgs boson production and decay rates.
The main effect of $\rho_{tt}$ is for $gg \to h$,
which is given by~\cite{Brod:2013cka, Nishiwaki:2013cma}
\begin{align}
\mu_{gg{\rm F}}
&\equiv \frac{\sigma(gg \to h)}{ \sigma(gg \to h)_{\rm SM}} \notag\\
&\simeq \bigl| (1.05 -0.08i)[ -\sin\gamma +\cos\gamma\,{\rm Re}(\rho_{tt})/\lambda_t]
 \bigr.\notag\\
&\quad \bigl. -0.05 +0.08i \bigr|^2
 +2.57 [\cos\gamma\,{\rm Im}(\rho_{tt})/\lambda_t]^2,\end{align}
through triangle loop diagram, where the absorptive,
i.e. explicit $i$ terms arise from light quark loops.
The $({\rm Re}\,\rho_{tt})^2$ effect is
suppressed by $(\cos\gamma)^2$, or alignment,
where we shall take the value of $\cos\gamma = 0.3$
(corresponding to $-\sin\gamma = 0.954$)
that may still be allowed by data~\cite{Altunkaynak:2015twa}.
Of interest is the $\sin\gamma\cos\gamma$ interference term
between the SM and extra Yukawa coupling,
which allows better sensitivity to ${\rm Re}\,\rho_{tt}$.
Although the $({\rm Im}\,\rho_{tt})^2$ term is
also suppressed by $(\cos\gamma)^2$,
but because the imaginary part of
the extra Yukawa coupling 
leads to a $\gamma_5$ coupling, the term receives
a numerical factor of order 2.6
 (see e.g. Ref.~\cite{Djouadi:2005gj} for
  discussion of pseudoscalar coupling),
making $gg \to h$ sensitive to larger values of $({\rm Im}\,\rho_{tt})^2$,
which links with EWBG.

The $h\to WW^*,\; ZZ^*$ rates are modified by the overall factor $\sin^2\gamma$
from the SM ones.
As for $h \to \gamma\gamma$ decay, which also arises from triangle loop,
the rate is only mildly affected~\cite{Brod:2013cka, Nishiwaki:2013cma}
by $\cos\gamma\, \rho_{tt}$
\begin{align}
\mu^{\gamma\gamma}
&\equiv \frac{\Gamma(h\to \gamma\gamma)}{\Gamma(h\to \gamma\gamma)_{\rm SM}}
\simeq \bigl| -(1.27+0.01i) \sin\gamma \bigr. \notag \\
 & \bigl. -0.28 [ -\sin\gamma
   +\cos\gamma\,{\rm Re}(\rho_{tt})/\lambda_t] +0.01-0.01 i \bigr|^2\notag\\
 & +0.18[\cos\gamma\,{\rm Im}(\rho_{tt})/\lambda_t]^2,
\end{align}
because of $W$ boson dominance in the loop.

\begin{figure*}[t!]
\center
\includegraphics[width=6cm]{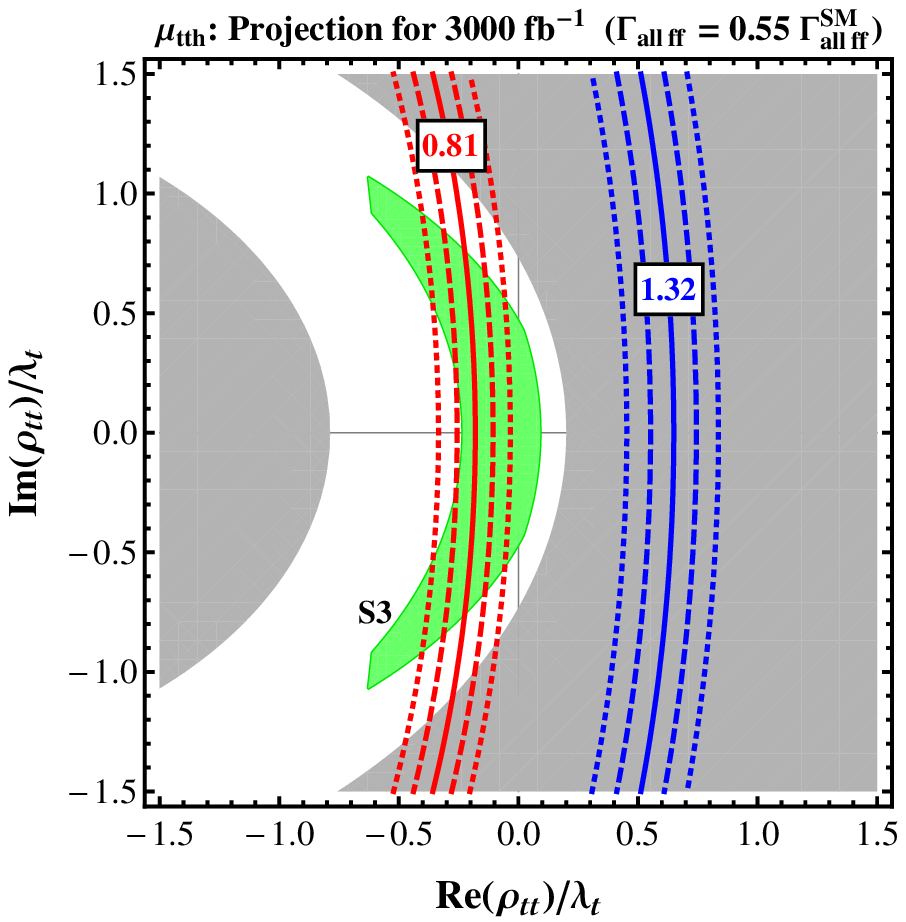} \hskip0.3cm
\includegraphics[width=6cm]{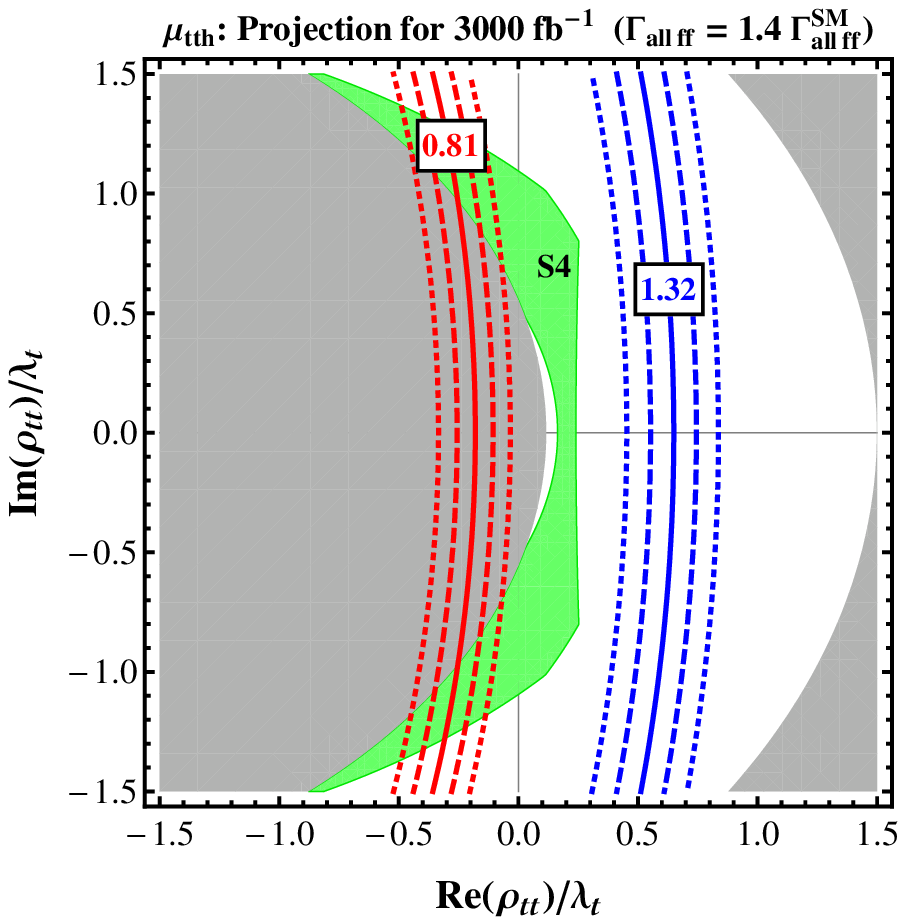} 
\caption{
Same as Fig.~\ref{fig:fig1} [right], but for $\Gamma_{{\rm all}~ff} =
  0.55 \Gamma_{{\rm all}~ff}^{\rm SM}$ [left],
 $1.4 \Gamma_{{\rm all}~ff}^{\rm SM}$ [right].
The green regions are for Scenarios 3 or 4 defined in
Eqs.~(\ref{eq:ggF-proj-S3}) and ~(\ref{eq:ggF-proj-S4}).
}
\label{fig:fig2}
\end{figure*}

Using the Run~1 combination of ATLAS and CMS results~\cite{Khachatryan:2016vau},
constraints on real and imaginary parts of $\rho_{tt}$ for
$\cos\gamma = 0.3$ with $\sin\gamma < 0$ are shown in Fig.~\ref{fig:fig1}.
We use the ten-parameter fit to the three decay channels
$h \to \gamma\gamma, ZZ^*, WW^*$ with $gg$F+$ttH$ or VBF+$VH$ production,
and individual $2\sigma$ constraints from the six signal strengths
are overlaid, resulting in the gray shaded regions.
The right-hand side (r.h.s.) comes from the $h\to WW^*$ decay mode,
while l.h.s. comes from $h \to ZZ^*$.
The actual sensitivity to ${\rm Re}\,\rho_{tt}$ and
${\rm Im}\,\rho_{tt}$ is mainly driven by $gg \to h$ production.

For the direct probe of $tth$ coupling,
we calculate the $pp \to t\bar t h$ cross section at 
leading order (LO) by Monte Carlo event generator
MadGraph5\_aMC@NLO~\cite{Alwall:2014hca}
with the parton distribution function set NN23LO1~\cite{Ball:2013hta}.
In particular, we use the Higgs Characterisation model~\cite{higgschar}
implemented in FeynRules~2.0~\cite{Alloul:2013bka} framework,
where model details can be found in Ref.~\cite{Demartin:2014fia}.
We ignored contributions other than from $tth$ coupling.
The signal strength for 13 TeV LHC
can be approximated by
\begin{align}
\mu_{t\bar t h}
&\simeq [ -\sin\gamma + \cos\gamma\,{\rm Re}(\rho_{tt})/\lambda_t ]^2\notag\\
&\quad +0.45[\cos\gamma\,{\rm Im}(\rho_{tt})/\lambda_t]^2, \quad \quad ({\rm 13~TeV})
\end{align}
with mild modification of the 0.45 coefficient to 0.46 for 14 TeV LHC.
Thus, the $({\rm Re}\,\rho_{tt})^2$ and $({\rm Im}\,\rho_{tt})^2$
terms are suppressed by $|\cos\gamma|^2$,
but the $\sin\gamma\cos\gamma$ interference term
brings in better sensitivity to ${\rm Re}\,\rho_{tt}$.
We take the ATLAS Run~2 update of $\mu_{t\bar t h}$,
Eq.~(\ref{eq:mu-ATL}), and display the $2\sigma$ allowed range
in Fig.~\ref{fig:fig1}[left] as marked.
Because of the mild excess, a positive ${\rm Re}\,\rho_{tt}$
is preferred, but partially excluded by the indirect data.
However, a broad range of $|{\rm Im}\,\rho_{tt}|$
is allowed, extending beyond 1 if the ${\rm Re}\,\rho_{tt}$
interference effect is destructive.

Thus, current data between indirect and direct probes
of $\rho_{tt}$ are quite consistent with
electroweak baryogenesis~\cite{Fuyuto:2017ewj} by 2HDM without NFC.

It is of interest, therefore, to project the reach for HL-LHC.
This is shown as the blue and red lines in Fig.~\ref{fig:fig1}[right]
for $\mu_{t\bar t h} = 0.81, 1.32$, corresponding to
$-2\sigma$ and central value in Eq.~(\ref{eq:mu-ATL}), respectively.
Considering the trend in measurements of $\mu_{t\bar t h}$,
we do not display the $+2\sigma$ case.
On the other hand, we take an optimistic 5\% as the
$1\sigma$ uncertainty reach for ultimate HL-LHC sensitivity,
based on current projections~\cite{ATLAS-Higgs-projection,CMS:2017cwx},
and anticipating a combination of ATLAS and CMS data.

The signal strength measurements for $gg$F 
and VBF 
production should also improve in the same time frame. 
With 3000 fb$^{-1}$, the projected uncertainties by ATLAS and
CMS~\cite{ATLAS-Higgs-projection,CMS:2017cwx}
are $\sim 5$\% for $h \to \gamma\gamma, ZZ^*$ with $gg$F production,
and $\sim 10$\% for $h \to \gamma\gamma, WW^*$ with VBF production.
In Fig.~\ref{fig:fig1}[right], HL-LHC projections for these four modes
are shown for the two following scenarios:
\begin{align}
&\mu_{gg{\rm F}}^{\gamma\gamma, ZZ} = 1,~
  \mu_{\rm VBF}^{\gamma\gamma, WW} = 1 \quad {\rm (Scenario~1)}, \notag\\
&\mu_{gg{\rm F}}^{\gamma\gamma, ZZ} = 1.2,~
  \mu_{\rm VBF}^{\gamma\gamma, WW} = 0.8 \quad {\rm (Scenario~2)}. \label{eq:ggF-proj}
\end{align}
Assuming $\pm 10$\% ($20$\%) ranges for $gg$F (VBF) modes as 2$\sigma$ uncertainties
with 3000 fb$^{-1}$ data, we overlay the four constraints
to obtain the green (denoted as S1) and yellow (S2) allowed regions.
We keep the Run 1 exclusion 
(gray shaded) regions for comparison.
In each scenario, the combination of the two $gg$F modes
gives the crescent-shaped allowed region, but
$h \to \gamma\gamma$ with VBF production cuts off 
part of the positive ${\rm Re}\, \rho_{tt}$ region in Scenario 1.
This is because $\cos\gamma = 0.3$ 
is a bit too large to sustain SM-like value for $\mu_{\rm VBF}^{\gamma\gamma}$,
which was assumed in Scenario 1.
Put differently, if $\cos\gamma$ is not too small,
we would have other means (e.g. combined fit) of 
probing it by the HL-LHC era.
We keep $\cos\gamma = 0.3$ for sake of comparison,
and to maximize the interference effect.

In Scenario 1, there is no overlap with the $\mu_{t\bar t h} \sim 1.32$ band,
while overlap with the $\mu_{t\bar t h} \sim 0.81$ band implies preference
for $|{\rm Im}\,\rho_{tt}| \sim 1$.
If the latter is found, it would provide insight on EWBG
without probing CPV directly.
On the other hand, Scenario 2 is compatible with the $\mu_{t\bar t h} \sim 1.32$ band,
which would imply a nonzero ${\rm Re}\,\rho_{tt}$ in the range
$0.4 \lesssim {\rm Re}\,\rho_{tt} \lesssim 0.8$,
with $|{\rm Im}\,\rho_{tt}| \lesssim 1.2$.

We have assumed the inclusive results for the $t\bar t h$ signal strength.
We note that, as data increases, separate production/decay channels would likely be disentangled.

\section{Simplified Effect of Light Fermions} 

But there is a catch in the discussion above.
Analogous to Eq.~(\ref{eq:tth}), the 2HDM without NFC
brings in extra Yukawa couplings that
modify $hbb$, $h\tau\tau$ and $hcc$ couplings.
While they give minor modifications to $gg \to h$ and $h \to \gamma\gamma$,
the major impact is on the total $h$ width, $\Gamma_h$,
which is not well measured yet.

As individual processes are also not yet well measured,
rather than several couplings,
we treat the partial width of $h \to$ light fermions as
the single overall effect, 
\begin{align}
\frac{\Gamma_{{\rm all}\;ff}}{\Gamma_{{\rm all}\;ff}^{\rm SM}}
= \frac{\Gamma_{bb} +\Gamma_{\tau \tau} +\Gamma_{cc} +\cdots }{\Gamma_{bb}^{\rm SM} +\Gamma_{\tau \tau}^{\rm SM} +\Gamma_{cc}^{\rm SM} +\cdots}. \label{eq:klf}
\end{align}
Taking branching ratio values~\cite{deFlorian:2016spz}
for $m_h = 125.09$ GeV, the total width is modified as
\begin{align}
\frac{\Gamma_h}{\Gamma_h^{\rm SM}}
&\simeq
0.67\frac{\Gamma_{{\rm all}\;ff}}{\Gamma_{{\rm all}\;ff}^{\rm SM}}
+0.24\sin^2\gamma +0.08\mu_{gg{\rm F}}.
\end{align}
We stress that 2HDM without NFC can relatively easily
lead to $\Gamma_h$ that differs from SM.

Similar to Fig.~\ref{fig:fig1}[right], we display the effect of
$\Gamma_{{\rm all}\;ff} = 0.55\ (1.4) \Gamma_{{\rm all}\;ff}^{\rm SM}$
in Fig.~\ref{fig:fig2}[left] ([right]), where the 
broad ``white crescent'' allowed by the Run 1 ATLAS-CMS combination
is moved leftward (rightward).
For $\Gamma_{{\rm all}\;ff} = 0.55\ \Gamma_{{\rm all}\;ff}^{\rm SM}$,
both the $gg$F and VBF signal strengths tend to be enhanced via
the enhancements in the $h$ decay branching ratios.
Hence, we tailor the following scenario for the HL-LHC projection:
\begin{align}
&\mu_{gg{\rm F}}^{\gamma\gamma, ZZ} = 1.2,~
  \mu_{\rm VBF}^{\gamma\gamma, WW} = 1.2 \quad {\rm (Scenario~3)}, \label{eq:ggF-proj-S3}
\end{align}
with $2\sigma$ range as in Scenarios 1 and 2,
shown as the green shaded allowed region in Fig.~\ref{fig:fig2}[left].
In this case, the crescent-shaped region is cut from left 
by $\mu_{\rm VBF}^{\gamma\gamma}$.
For $\Gamma_{{\rm all}\;ff} = 1.4\ \Gamma_{{\rm all}\;ff}^{\rm SM}$,
the effect for the signal strengths is opposite and, hence, we introduce
\begin{align}
&\mu_{gg{\rm F}}^{\gamma\gamma, ZZ} = 0.8,~
  \mu_{\rm VBF}^{\gamma\gamma, WW} = 0.8 \quad {\rm (Scenario~4)}. \label{eq:ggF-proj-S4}
\end{align}
The $2\sigma$ allowed region, shown as the green region in Fig.~\ref{fig:fig2}[right],
is similar to S1 in Fig.~\ref{fig:fig1}[right].

The overlap of the S3 region and
$\mu_{t\bar t h} \sim 0.81$ band in Fig.~\ref{fig:fig2}[left]
illustrates the situation where ${\rm Re}\,\rho_{tt}$ is slightly
negative, with $|{\rm Im}\,\rho_{tt}|$ up to 1 fully allowed.
The overlap of the S4 region and $\mu_{t\bar t h} \sim 0.81$ band
in Fig.~\ref{fig:fig2}[right]
illustrates the situation where 
$|{\rm Im}\,\rho_{tt}| \sim 1$.
The $\mu_{t\bar t h} \sim 1.32$ band illustrates 
a tension between indirect and direct probes of $\rho_{tt}$
for both S3 and S4.

What Fig.~\ref{fig:fig2} really 
illustrates is the need for 
a better measurement of $\Gamma_h$~\cite{tth-Gam_h}.
With current projections at 50\% of
$\Gamma_h^{\rm SM}$~\cite{ATL-15} for HL-LHC,
we can only hope that it can be improved with actual data.
The 2HDM without NFC
would shift all diagonal Yukawa couplings,
in principle by same order as the corresponding
SM Yukawa coupling, modulated by $\cos\gamma$.
More precise measurements of
$\mu_{c\bar c}$, $\mu_{\tau^+\tau^-}$ and especially
$\mu_{b\bar b}$ may help.
Another approach, for example, is VBF production followed by $h \to VV^*$.
Measuring $\mu_{\rm VBF}^{VV^*}$ could probe $\Gamma_h$,
with only mild dependence on a small $\cos\gamma$.


\section{Discussion and Conclusion} 

We stress that 2HDM without NFC offer new Yukawa couplings
that could alter all $ffh$ couplings from SM values,
modulated by $\cos\gamma$.
This makes clear the importance of
a complete program to measure $\mu_{b\bar b}$, $\mu_{\tau^+\tau^-}$,
$\mu_{\mu^+\mu^-}$, and even $\mu_{c\bar c}$
if charm tagging could be vastly improved.

A second point to note is that each one of these
diagonal Yukawa coupling corrections are generally complex.
For example, ${\rm Im}(\rho_{tt})$ contributes to
the electron EDM through two-loop 
contributions. Under the assumption that
the electron Yukawa coupling is SM-like ($\rho_{ee}=0$)
and heavy scalar contributions are negligible,
the recent ACME result~\cite{Baron:2013eja} would imply
$|\cos\gamma\,{\rm Im}(\rho_{tt})/\lambda_t|
 < 0.01$~\cite{Fuyuto:2017ewj, Brod:2013cka}.
However, allowing for a complex, in particular imaginary,
$\rho_{ee}$ with strength similar to $\lambda_e$ of SM,
it can in principle induce cancellation~\cite{Fuyuto:2017ewj}
of the two-loop effect.
We had tacitly assumed this in exploring $t\bar th$,
and illustrates how the 2HDM without NFC could affect flavor physics.
Note that, given that $\lambda_t \cong 1$ is already known,
the dominant eigenvalue of the other combination of
the two $u$-type Yukawa matrices, viz. $\rho_{tt}$,
is likely ${\cal O}(\lambda_t)$ hence ${\cal O}(1)$,
and with phase arbitrary.
Similar argument would hold for $\rho_{bb}$ and $\rho_{\tau\tau}$.

In contrast to usual effective Lagrangian discussions,
the interaction terms reflected in Eqs.~(\ref{eq:tth}) and (\ref{eq:tch})
are \emph{as fundamental as} the Yukawa couplings in SM.
On one hand, they are well hidden by approximate alignment,
or the smallness of $\cos\gamma$.
Thus, the direct $pp \to t\bar th$ and other indirect probes
would rapidly weaken for smaller $\cos\gamma$.
Furthermore, as outlined in Ref.~\cite{Hou:2017hiw},
the mass-mixing hierarchy, or some flavor-organization principle,
could control FCNH involving lighter generations,
and together with approximate alignment, can fully replace the NFC condition
to explain the absence of low energy FCNH effects.
But alignment need not~\cite{Hou:2017hiw} imply decoupling,
and the exotic Higgs sector could well be sub-TeV in mass.
If alignment is effective, one would have to probe this
exotic Higgs sector, for example via $cg \to tH^0$, $tA^0$ production,
leading to novel $tt\bar c$ (same-sign top) and $tt\bar t$ (triple top)
signatures~\cite{Kohda:2017fkn} at the LHC.

With the hope that alignment does not work too well,
i.e. $\cos\gamma$ is not overly small, what is the future outlook?
First, $\cos\gamma$ should be studied more generally,
free of the $Z_2$ symmetry mindset
 (i.e. beyond Ref.~\cite{cos_b-a}).
A measurement of, rather than constraint on, $\cos\gamma$
would be astounding.
Second, improved projections for HL-LHC is expected
with a CERN Yellow Report that is under preparation for
the European Particle Physics Strategy Update.
But the projections must be continuously updated
as experience is gained with larger datasets,
including on $\Gamma_h$ measurement.
Third, we are at the juncture of ILC(250)~\cite{Evans:2017rvt}
or CLIC(380)~\cite{CLIC:2016zwp} decision.
Although these machines are still far away,
they provide great hope for much more precision in
Higgs property measurements, which would provide
better indirect constraints, including on $\delta\Gamma_h$.
However, $t\bar th$ production would require
at least 500 GeV $e^+e^-$ collision energy.

Thus, the high energy extension of LHC looks more promising
for the nearer future on direct $t\bar th$ probe.
There is no doubt that a 100 TeV machine~\cite{mlm_bo},
though much farther away,
would advance the $t\bar th$ and exotic Higgs frontiers
by great stride.
If the alignment phenomenon reflects~\cite{Hou:2017hiw} ${\cal O}(1)$
Higgs quartic couplings within 2HDM,
there is likely another layer of BSM physics
at the 10 TeV scale to be explored.
Finally, we have advocated simple
probes of just measuring rates.
More sophisticated angular or asymmetry analyses~\cite{CPV-tth}
can probe the CPV nature of the $tth$ coupling directly.

%

In conclusion, the observation of $pp \to t\bar th^0$
at LHC is the first direct measurement of the top Yukawa coupling,
and offers a window on the extra Yukawa coupling $\rho_{tt}$
from a second Higgs doublet where the NFC condition is not imposed.
The large $\lambda_t \cong 1$ of SM provides the lever arm
to probe $\cos\gamma\,\rho_{tt}$ through interference,
where $\cos\gamma$ is the $CP$-even Higgs mixing angle.
The parameter space for electroweak baryogenesis offered by this 2HDM
can be probed by $CP$-conserving $t\bar th^0$ production rate
and Higgs property measurements.
The Achilles heel for this program at the HL-LHC
is the knowledge of the $h^0$ width, $\Gamma_{h^0}$,
and ATLAS and CMS should put a premium
on its improved measurement.

\vskip0.2cm
\noindent{\bf Acknowledgments} \
We thank K.-F. Chen and S. Paganis for discussions.
This research is supported by grants MOST 104-2112-M-002-017-MY3,
107-2811-M-002-039, and 106-2811-M-002-187.



\begin{thebibliography}{99}

%
\bibitem{Sirunyan:2018hoz}
  A.M.~Sirunyan {\it et al.} [CMS Collaboration],
  Phys.\ Rev.\ Lett.\  {\bf 120}, 231801 (2018).
%
\bibitem{Aaboud:2018urx}
  M.~Aaboud {\it et al.} [ATLAS Collaboration],
  Phys.\ Lett.\ B {\bf 784}, 173 (2018).
%
\bibitem{Aaboud:2017jvq}
  M.~Aaboud {\it et al.} [ATLAS Collaboration],
  Phys.\ Rev.\ D {\bf 97}, 072003 (2018).
%
\bibitem{Glashow:1976nt}
  S.L.~Glashow, S.~Weinberg,
  Phys.\ Rev.\ D {\bf 15}, 1958 (1977).
%
\bibitem{Cheng:1987rs}
  T.-P.~Cheng, M.~Sher,
  Phys.\ Rev.\ D {\bf 35}, 3484 (1987).
%
\bibitem{Hou:1991un}
  W.-S.~Hou,
  Phys.\ Lett.\ B {\bf 296}, 179 (1992).
%
\bibitem{Chen:2013qta}
  K.-F.~Chen {\it et al.}, 
  Phys.\ Lett.\ B {\bf 725}, 378 (2013).
%
\bibitem{rhoui}
 Note that the $\rho_{ui}$, $\rho_{iu}$ elements ($i = 1$--3)
 would easily generate large effects, hence would have to be
 rather small. This is consistent with the mass-mixing pattern.
%
\bibitem{Altunkaynak:2015twa}
  B.~Altunkaynak {\it et al.},
  Phys.\ Lett.\ B {\bf 751}, 135 (2015).

%
\bibitem{cos_b-a}
  See e.g. D.~Chowdhury, O.~Eberhardt,
  JHEP {\bf 1805}, 161 (2018);
  J.~Haller {\it et al.}, 
  Eur.\ Phys.\ J.\ C {\bf 78}, 675 (2018).
%
\bibitem{Khachatryan:2016vau}
  G.~Aad {\it et al.} [ATLAS and CMS Collaborations],
  JHEP {\bf 1608}, 045 (2016).
%
\bibitem{Hou:2017hiw}
  W.-S.~Hou, M.~Kikuchi,
  EPL {\bf 123}, 11001 (2018).
%
\bibitem{Fuyuto:2017ewj}
  K.~Fuyuto, W.-S.~Hou, E.~Senaha,
  Phys.\ Lett.\ B {\bf 776}, 402 (2018).
  This paper discusses further processes that
  are suppressed, or not, by $\cos\gamma$.
%
\bibitem{Kohda:2017fkn}
  M.~Kohda, T.~Modak, W.-S.~Hou,
  Phys.\ Lett.\ B {\bf 776}, 379 (2018).
%
\bibitem{tch-13}
  For updates based on 13 TeV data, see
  M.~Aaboud {\it et al.} [ATLAS Collaboration],
  JHEP {\bf 1710}, 129 (2017),
%
  and 
  Phys.\ Rev.\ D {\bf 98}, 032002 (2018);
  A.M.~Sirunyan {\it et al.} [CMS Collaboration],
  JHEP {\bf 1806}, 102 (2018).
%
\bibitem{Brod:2013cka}
  J.~Brod, U.~Haisch, J.~Zupan,
  JHEP {\bf 1311}, 180 (2013).
%
\bibitem{Nishiwaki:2013cma}
  K.~Nishiwaki, S.~Niyogi, A.~Shivaji,
  JHEP {\bf 1404}, 011 (2014).
%
\bibitem{Djouadi:2005gj}
  A.~Djouadi,
  Phys.\ Rept.\  {\bf 459}, 1 (2008).
%
\bibitem{Alwall:2014hca}
  J.~Alwall {\it et al.},
  JHEP {\bf 1407}, 079 (2014).
%
\bibitem{Ball:2013hta}
  R.D.~Ball {\it et al.} [NNPDF Collaboration],
  Nucl.\ Phys.\ B {\bf 877}, 290 (2013).
%
\bibitem{higgschar}
 The Higgs Characterisation model
 \url{http://feynrules.irmp.ucl.ac.be/wiki/HiggsCharacterisation}
%
\bibitem{Alloul:2013bka}
  A.~Alloul {\it et al.}, 
  Comput.\ Phys.\ Commun.\  {\bf 185}, 2250 (2014).
%
\bibitem{Demartin:2014fia}
  F.~Demartin {\it et al.}, 
  Eur.\ Phys.\ J.\ C {\bf 74}, 3065 (2014).
%
\bibitem{ATLAS-Higgs-projection}
ATLAS Collaboration,
  ATL-PHYS-PUB-2014-016.
%
\bibitem{CMS:2017cwx}
  CMS Collaboration,
  CMS-PAS-FTR-16-002.
%
%
\bibitem{deFlorian:2016spz}
  D.~de Florian {\it et al.} [LHC Higgs Cross Section Working Group],
  arXiv:1610.07922 [hep-ph].
%
\bibitem{tth-Gam_h}
  While our two bands for $\mu_{t\bar th}$ just
  illustrate the span from below to above 1 for final HL-LHC values,
  we note that $\mu_{t\bar th}$ measurement itself
  would depend also on $\Gamma_h$.
%
\bibitem{ATL-15}
  ATLAS Collaboration,
  ATL-PHYS-PUB-2015-024.
%
\bibitem{Baron:2013eja}
  J.~Baron {\it et al.} [ACME Collaboration],
  Science {\bf 343}, 269 (2014).
%
\bibitem{Evans:2017rvt}
 L.~Evans, S. Michizono (Ed.) [Linear Collide Collaboration],
  arXiv:1711.00568 [physics.acc-ph].
%
\bibitem{CLIC:2016zwp}
  M.J.~Boland {\it et al.} [CLIC and CLICdp Collaborations],
  arXiv:1608.07537 [physics.acc-ph].
%
\bibitem{mlm_bo}
  For beyond HL-LHC hadron collider physics,
  see e.g. the plenary talk by M.L. Mangano
  at Large Hadron Collider Physics conference, June 2018, Bologna, Italy.
%
\bibitem{CPV-tth}
For a non-exhaustive list, see e.g.
  J.~Ellis {\it et al.}, 
  JHEP {\bf 1404}, 004 (2014);
  F.~Boudjema {\it et al.}, 
  Phys.\ Rev.\ D {\bf 92}, 015019 (2015);
  M.R.~Buckley, D.~Gon\c{c}alves,
  Phys.\ Rev.\ Lett.\  {\bf 116}, 091801 (2016);
  H.-L.~Li {\it et al.}, 
  Chin.\ Phys.\ C {\bf 40}, 063102 (2016);
  N.~Mileo {\it et al.}, 
  JHEP {\bf 1607}, 056 (2016);
  A.V.~Gritsan {\it et al.}, 
  Phys.\ Rev.\ D {\bf 94}, 055023 (2016);
  J.~Li {\it et al.}, 
  Phys.\ Lett.\ B {\bf 779}, 72 (2018);
  D.~Gon\c{c}alves, K.~Kong, J.H.~Kim,
  JHEP {\bf 1806}, 079 (2018).
  and references therein.

\end{thebibliography}
\end{document}